\documentclass[reqno,aps,prd,twocolumn,showpacs,showkeys,preprintnumbers,amsmath,amssymb,amsfonts]{revtex4}

\usepackage[dvips]{graphicx}
\usepackage{dcolumn}
\usepackage{bm}
\usepackage[bookmarksnumbered]{hyperref}

\numberwithin{equation}{section}

\begin{document}

\preprint{GW4.31}

\pdfbookmark{Orbital period derivative of a binary system}{tit}
\title{Orbital period derivative of a binary system using an exact orbital energy equation}


\author{Vikram H. Zaveri}
\email{cons\_eng1@yahoo.com}
\affiliation{B-4/6, Avanti Apt., Harbanslal Marg, Sion, Mumbai 400022 INDIA}


\date{March 17, 2010}

\begin{abstract}
It is proposed that the equations of motion in periodic relativity which yielded major predictions of general relativity are exact in nature and can be applied to pulsars and inspiraling compact binaries for analyzing orbital period derivative and two polarization gravitational wave forms. Exactness of these equations eliminates the need for higher order $x$PN corrections to the orbital energy part of the balance equation. This is mainly due to the introduction of dynamic WEP which states that the gravitational mass is equal to the relativistic mass.
\end{abstract}


\maketitle

\section{Introduction}

The famous quadrupole formalism of Einstein \cite{9} is the lowest order wave generation formalism in the Newtonian limit $1/c \rightarrow 0$. Using this formalism, Peters and Mathews \cite{10} calculated the time average of the energy radiated from a system of two gravitating masses. From these results, secular decays of the semi major axis, and eccentricity were found by Peters \cite{4}. Based on this work, Esposito and Harrison \cite{11}, and Wagoner \cite{12} heuristically formulated the orbital period derivative for the binary pulsar system in light of the discovery of a radio pulsar PSR 1913+16 by Hulse and Taylor \cite{13}. Timing measurement of PSR 1913+16 by Weisberg and Taylor \cite{14} provided confirmation for the rate of orbital period decay due to gravitational radiation damping as predicted by general relativity. Subsequently both the theoretical model and the measurement accuracies were improved over a period of time \cite{8,15,16,17,18,19,20,21,29,30}.

In periodic relativity theory (PR) \cite{1,33}, we extended the concept of space-time symmetry to include heavier de Broglie particles, which enabled us to introduce energy-momentum invariant into the space-time invariant resulting in an invariant relationship between the force and the energy. We utilized this relation to address two-body problem in gravitational field by proposing the following equation of orbital motion  
\begin{align}\label{1.1}
\frac{d}{dt}(E-m_0c^2)=-\frac{\mu m}{r^2}\mathbf{\hat r}\frac{d\mathbf{r}}{dt}\left(\cos{\psi}+\sin{\psi}\right),
\end{align}
where $m=\gamma m_0$ is the relativistic mass, $E$ total energy of the particle, $\psi$ is the angle between the radial vector $\mathbf{r}$ and the velocity vector $\mathbf{v}$.
Factor $\left(\cos{\psi}+\sin{\psi}\right)$ is responsible for giving us geodesic like trajectories. Appearance of this factor is due to the introduction of the Newtonian gavitational potential in the form of summation of two components due to acceleration 
$(\mu m/r^2)\cos{\psi}$ and $(\mu m/r^2)\sin{\psi}$ acting on the gravitating body. With this set-up for the equation of motion, we successfully obtained exactly the same lowest order expressions for gravitational redshift and deflection of light as in general relativity. We also obtained second order non-homogeneous non-linear differential equation of motion equivalent to general relativity which yielded same values of the perihelic precession for the planets of the solar system. These are reasons enough to conclude that Eq.~\eqref{1.1} is an exact equation of motion \cite{2} and can provide us exact orbital energy for two body systems. If we indiscriminately introduce $x$PN corrections to these equations, there is a good chance that we may not be able to derive correct formula for gravitational redshift and bending of light from such corrected equations. This means that we have locked on to an orbital energy equation which we cannot alter. This is mainly because the present theory is energy based theory. In comparision, general relativity uses Newtonian orbital energy equations and adds higher order post-Newtonian corrections \cite{22,23,24,25,27,28,6,8} to get desired relativistic effects for computing center of mass binding energy and gravitational wave energy flux. In PR, these corrections are fundamentally built into the orbital energy equation. The basis of this equation is the introduction of dynamic WEP(weak equivalence principle) in PR. Dynamic WEP states that the gravitational mass is equal to the relativistic mass. For more details see \cite{1,33}. This situation provides an ideal testing ground for proving this theory in the area of gravitational radiation.

\section{Orbital Energy of a Binary System}
For a binary system, we can rewrite Eq.~\eqref{1.1} as 
\begin{align}\label{2.1}
\frac{d}{dt}(E-m_2c^2)=-\frac{\mu_1 m_2 \gamma}{r^2}\mathbf{v} \left(\cos{\psi}+\sin{\psi}\right)\mathbf{\hat r},
\end{align}
where $m_1$ is the mass of the pulsar and $m_2$ that of the companion star. For Binary, $\mu_1=Gm_1$, $r$ is the separation. Total orbital energy $E^0$ of the system can be given by
\begin{align}\label{2.2}
E^0=\left(E-m_2c^2\right)+\int \frac{\mu_1 m_2 \gamma}{r^2} \frac{dr}{dt} \left(\cos{\psi}+\sin{\psi}\right)dt.
\end{align}
If we evaluate $E^0$ at periastron, we can intorduce following approximations.
$\psi=\pi/2$ and $m_2\gamma=constant$. Therefore,
\begin{align}
E^0=\left(E-m_2c^2\right)+ \mu_1 m_2 \gamma \left(-\frac{1}{r}\right),\label{2.3}\\ 
E^0=m_2c^2\left[\left(\gamma-1\right)-\frac{\mu_1\gamma}{rc^2}\right],\label{2.4}\\
E^0=m_2c^2\left[\left(1-\frac{v_p^2}{c^2}\right)^{-\frac{1}{2}}\left
\{1-\frac{\mu_1}{rc^2}\right\}-1\right],\label{2.5}
\end{align}
where $v_p$ is the velocity at periastron. One can observe that the $x$PN corrections are naturally and fundamentally built into the orbital energy equation Eq.~\eqref{2.5}. Unlike general relativity, there is no limit to which this equation can be expanded. However, here we will expand Eq.~\eqref{2.5} equivalent to 3.5PN order in general relativity.
\begin{align}\label{2.6}
 \begin{split}
E^0=m_2c^2\biggl[&\left(1+\frac{1}{2}\frac{v_p^2}{c^2}+\frac{3}{8}\frac{v_p^4}{c^4}+\frac{5}{16}\frac{v_p^6}{c^6}+\frac{35}{128}\frac{v_p^8}{c^8}+\ldots\right) \\
&\times\left\{1-\frac{\mu_1}{rc^2}\right\}-1\biggr],
 \end{split}
\end{align}
\vspace{-12mm}
\begin{align}\label{2.7}
 \begin{split}
E^0=-\biggl[&\frac{\mu_1 m_2}{2a}+\frac{1}{8}\frac{\mu_1^2 m_2(1+e)(1-3e)}{c^2 a^2 (1-e)^2}\\
&+\frac{1}{16}\frac{\mu_1^3 m_2 (1+e)^2(1-5e)}{c^4a^3(1-e)^3}\\
&+\frac{5}{128}\frac{\mu_1^4m_2(1+e)^3(1-7e)}{c^6a^4(1-e)^4}+\mathcal{O}\left(\frac{1}{c^8}\right)\biggr],
 \end{split}
\end{align}
where $v_p=h/r_p$, $h^2=\mu_1 a(1-e^2)$ and $r_p=a(1-e)$ at periastron.
Variation of $E^0$ with time is then given by Eq.~\eqref{2.8}. It is to be noted here that introduction of the deviation to flat Minkowski metric \cite{1,33} in Eq.~\eqref{2.2} and Eq.~\eqref{2.8} does not alter the formulation because the effect gets cancelled out.
\begin{widetext}
\begin{align}\label{2.8}
 \begin{split}
\frac{dE^0}{dt}&=\biggl[\frac{\mu_1 m_2}{2a^2}-\frac{1}{4}\frac{\mu_1^2 m_2(1+e)(3e-1)}{c^2 a^3 (1-e)^2}-\frac{3}{16}\frac{\mu_1^3 m_2 (1+e)^2(5e-1)}{c^4a^4(1-e)^3}-\frac{5}{32}\frac{\mu_1^4m_2(1+e)^3(7e-1)}{c^6a^5(1-e)^4}\biggr]\frac{da}{dt}\\
&+\biggl[\frac{\mu_1^2 m_2 e}{c^2 a^2 (1-e)^3}+\frac{1}{8}\frac{\mu_1^3 m_2 (7e^2+4e+1)}{c^4a^3(1-e)^4}+\frac{15}{8}\frac{\mu_1^4m_2 e(1+e)^2}{c^6a^4(1-e)^5}\biggr]\frac{de}{dt} +\mathcal{O}\left(\frac{1}{c^8}\right).
 \end{split}
\end{align}
\end{widetext}

\section{Orbital Period Derivative}

Laser interferometric observations of gravitational wa\-ves demands energy balance equation computed to an extraordinarily high degree of accuracy of order $1/c^6$. Equations of variation with time of the orbital frequency and orbital phase of an inspiralling compact binary are derived from this energy balance equation and the theoretical templates of the compact inspiral binary is obtained by introducing these highly accurate parameters into binary's two polarization wave-forms $h_+$ and $h_{\times}$ \cite{6}. In case of binary pulsars, the energy balance equation is given by
\begin{align}\label{3.1}
\frac{dE^0}{dt}=-\mathcal{L},
\end{align}
where $E^0$ is the orbital energy and $\mathcal{L}$ the total gravitational luminosity (or wave flux) of the source. The time average value $\langle\mathcal{L}\rangle$ has been computed to 1PN order by Blanchet and Sch$\ddot{a}$fer\cite{3} and can be given in our terminology by
\begin{align}\label{3.2}
 \begin{split}
\langle\mathcal{L}\rangle&=\frac{1024}{5Gc^5}\frac{\nu^2(-E^0)^5}{\mu^5(1-e^2)^{\frac{7}{2}}}\biggl\{1+\frac{73}{24}e^2+\frac{37}{96}e^4\\
&+\frac{-E^0}{168\mu c^2(1-e^2)}\biggl[13-6414e^2-\frac{27405}{4}e^4-\frac{5377}{16}e^6\\
&+\left(-840-\frac{6419}{2}e^2-\frac{5103}{8}e^4+\frac{259}{8}e^6\right)\nu \biggr] \biggr\},
 \end{split}
\end{align}
where $\mu=(m_1m_2)/m$, $~m=(m_1+m_2)$, $~\nu=\mu/m$ and $E^0$ to 1PN accuracy can be obtained from Eq.~\eqref{2.7} and given by
\begin{align}\label{3.3}
E^0&=-\frac{\mu_1 m_2}{2a}\biggl[1+\frac{1}{4}\frac{\mu_1 (1+e)(1-3e)}{a c^2 (1-e)^2}\biggr].
\end{align}
We can truncate Eq.~\eqref{2.8} to 1PN accuracy as follows.
\begin{align}\label{3.4}
\begin{split}
\frac{dE^0}{dt}&=\biggl[\frac{\mu_1 m_2}{2a^2}-\frac{1}{4}\frac{\mu_1^2 m_2(1+e)(3e-1)}{c^2 a^3 (1-e)^2}\biggr]\frac{da}{dt}\\
&+\biggl[\frac{\mu_1^2 m_2 e}{c^2 a^2 (1-e)^3}\biggr]\frac{de}{dt}.
\end{split} 
\end{align}
We will utilize following lowest order expression for $\langle de/dt\rangle$ given by Peters \cite{4}. 1PN expression could be more consistent but it may not affect the result significantly.
\begin{align}\label{3.5}
 \begin{split}
\left\langle\frac{de}{dt}\right\rangle=\frac{19e(1-e^2)(1+(121/304)e^2)}{12a(1+(73/24)e^2+(37/96)e^4)}\left\langle\frac{da}{dt}\right\rangle.
 \end{split}
\end{align}
We get following relation from Kepler's third law,
\begin{align}\label{3.6}
\frac{da}{dt}=\frac{2a}{3}\frac{\dot{P_b}}{P_b},
\end{align}
where $P_b$ is the orbital period and $\dot{P_b}$ orbital period derivative. Substitution of Eqs.~\eqref{3.5} and \eqref{3.6} in Eq.~\eqref{3.4}	yields,
\begin{align}\label{3.7}
\left\langle\frac{dE^0}{dt}\right\rangle=\frac{\mu_1m_2}{3a}\left[1-\frac{\mu_1(\sigma_1+\sigma_2)}{\sigma_3}\right]\frac{\dot{P_b}}{P_b},
\end{align}
\begin{align*}
\text{where}\qquad
\sigma_1=(-3+6e+(151/8)e^2+(149/4)e^3),\\
\sigma_2=((1081/32)e^4+(79/8)e^5+(111/32)e^6),\\
\sigma_3=6ac^2(1-e)^2(1+(73/24)e^2+(37/96)e^4).
\end{align*}
Introducing Eqs.~\eqref{3.2} and \eqref{3.7} into the balance Eq.~\eqref{3.1} and replacing $a$ with Kepler's third law in the form
\begin{align}\label{3.8}
\frac{1}{a}=\left(\frac{2\pi}{P_b}\right)^{\frac{2}{3}}\mu_2^{-\frac{1}{3}},\qquad\text{where}\quad \mu_2=G(m_1+m_2),
\end{align}
\begin{equation}\label{3.9}
\text{we get,}\qquad\quad\qquad\dot{P_b}=k_1(1+k_2),\qquad \text{where,}
\end{equation}
\begin{align}\label{3.10}
\begin{split}
k_1=&-\frac{192\pi}{5c^5}\left(\frac{2\pi G}{P_b}\right)^{\frac{5}{3}}\frac{m_1m_2(m_1+m_2)^{-\frac{1}{3}}}{(1-e^2)^{\frac{7}{2}}}\\&\times\left\{1+\frac{73}{24}e^2+\frac{37}{96}e^4\right\},
\end{split}
\end{align}
\begin{align}\label{3.11}
k_2=\left[(k_6-1)+\left(\frac{k_3k_4k_5k_6}{k_1}\right)\right],
\end{align}
\begin{align}\label{3.12}
 \begin{split}
k_6&=\left[1+\frac{1}{4}\frac{\mu_1 (1+e)(1-3e)}{a c^2 (1-e)^2}\right]^5\left[1-\frac{\mu_1(\sigma_1+\sigma_2)}{\sigma_3}\right]^{-1}\\
&=\left[1+\frac{5}{4}\frac{(1+e)(1-3e)}{c^2(1-e)^2}\left(\frac{2\pi G}{P_b}\right)^{\frac{2}{3}}m_1(m_1+m_2)^{-\frac{1}{3}}\right]\\
&\times\biggl[1+\frac{(\sigma_1+\sigma_2)m_1(m_1+m_2)^{-\frac{1}{3}}}{6c^2(1-e)^2(1+(73/24)e^2+(37/96)e^4)}\\
&\times\left(\frac{2\pi G}{P_b}\right)^{\frac{2}{3}}\biggr],
 \end{split}
\end{align}
\begin{align}\label{3.13}
k_3=-\frac{4\pi}{35}\left[\frac{m_1m_2(m_1+m_2)^{\frac{1}{3}}}{c^7(1-e^2)^{\frac{9}{2}}}\left(\frac{2\pi G}{P_b}\right)^{\frac{7}{3}}\right],
\end{align}
\begin{align}\label{3.14}
k_4=\left[1+\frac{1}{4}\frac{(1+e)(1-3e)}{c^2 (1-e)^2}\left(\frac{2\pi G}{P_b}\right)^{\frac{2}{3}}m_1(m_1+m_2)^{-\frac{1}{3}}\right],
\end{align}
\begin{align}\label{3.15}
\begin{split}
k_5=\biggl[&13-6414e^2-\frac{27405}{4}e^4-\frac{5377}{16}e^6
+\biggl(-840 \\
&-\frac{6419}{2}e^2-\frac{5103}{8}e^4+\frac{259}{8}e^6\biggr)\nu\biggr].
\end{split}
\end{align}
For binary pulsar PSR 1913+16, factor $k_2$ given by Eq.~\eqref{3.11} in this theory turns out to be $k_2=-1.865\times10^{-5}$ compared to $+2.15\times10^{-5}$ given by Blanchet and Sch$\ddot{a}$fer \cite{3}, which is $-1.1528$ times factor $k_2$ in our theory and $-60$ times the result of another closest rival theory of Spyrou and Papadopoulos \cite{31}. Therefore, in a way this theory is in remarkable agreement with that of Blanchet and Sch$\ddot{a}$fer. As of today this value remains below the accuracy in the measurement of $\dot{P_b}$ \cite{8,29,30}. The slight difference between the two theories is due to the different methods employed in obtaining the orbital energy equations. Eq.~\eqref{3.3} in present theory and Eq.~(3.35) in the theory of Blanchet and Sch$\ddot{a}$fer \cite{3}. Present theory uses basic mathematical tools involving very few approximations for deriving the exact orbital energy equation, whereas Blanchet and Sch$\ddot{a}$fer uses advanced mathematical tools involving plenty of approximations. These methods of post-Newtonian wave generation formalism are described in Blanchet and Damour \cite{32}, Damour and Deruelle \cite{26,15}. Again the stress energy tensor used by Blanchet and Sch$\ddot{a}$fer) uses Newtonian gravitational potential and classical kinetic energy where as present theory uses relativisitic gravitational potential and relativistic kinetic energy. The success of this theory is primarily due to the introduction of the relativistic mass of the orbiting body in the Newtonian theory of gravitation. This in turn provides further justification for the use of relativistic mass. In PR \cite{1,33} static WEP is modified to dynamic WEP which states that the gravitational mass is equal to the relativistic mass.

\section{Orbital Phase of an Inspiraling Compact Binary}

In this section, we will simply relate the orbital energy Eq.~\eqref{2.7} to the 3.5PN accurate gravitational wave-form model discussed by Blanchet et al.\cite{5}. Since the orbits of the inspiralling compact binaries are circularized, for $e=0$, Eq.~\eqref{2.7} reduces to 
\begin{align}\label{4.1}
E^0=-\frac{\mu_1 m_2}{2a}\biggl[1+\frac{1}{4}\frac{\mu_1}{ac^2} +\frac{1}{8}\frac{\mu_1^2}{a^2c^4}
+\frac{5}{64}\frac{\mu_1^3}{a^3c^6}\biggr]+\mathcal{O}\left(\frac{1}{c^8}\right).
\end{align}
We have the post-Newtonian parameter $\gamma$ given by, 
\begin{align}\label{4.2}
\gamma=\frac{Gm}{r_{12}c^2}=\frac{G(m_1+m_2)}{ac^2}=\frac{\mu_2}{ac^2}.
\end{align}
Substitution of Eq.~\eqref{4.2} in Eq.~\eqref{4.1} yields,
\begin{align}\label{4.3}
\begin{split}
E^0=-\frac{\mu c^2 \gamma}{2}\biggl[&1+\frac{1}{4}\left(\frac{\mu}{m_2}\right)\gamma +\frac{1}{8}\left(\frac{\mu}{m_2}\right)^2\gamma^2\\
&+\frac{5}{64}\left(\frac{\mu}{m_2}\right)^3\gamma^3\biggr]+\mathcal{O}\left(\frac{1}{c^8}\right).
\end{split}
\end{align}
Equation Eq.~\eqref{4.3} is comparable to those given in \cite{5,7}. The rest of the procedure for computing the orbital phase $\phi$ in terms of the frequency related parameter $x$ can be same as described in \cite{5,6}. One can either use the same 3.5PN accurate expression for the gravitational wave flux $\mathcal{L}$ \cite{5,6} or reevaluate $\mathcal{L}$ in consideration of the new equations of motion resulting from Eq.~\eqref{1.1}.

\section{Conclusion}
We have derived precise value for the orbital period derivative which is very close to the currently accepted theoretical value derived by Blanchet and Sch$\ddot{a}$fer \cite{3}. This needs to be experimentally verified for establishing the correct theoretical approach. It would be impossible to derive the expression for the orbital period derivative in this theory without the use of relativistic mass for the orbiting body in the inverse square law of gravitation. In PR \cite{1,33} static WEP is modified to dynamic WEP which states that the gravitational mass is equal to the relativistic mass. It is the use of relativistic mass that eliminates the need for higher order $x$PN corrections to the orbital energy part of the balance equation. Further justification for using the relativistic mass is discussed at length in \cite{1,33}.

\pdfbookmark{References}{Ref}
\bibliographystyle{amsalpha}

\end{document}